\documentclass[10pt,aps,superscriptaddress,twocolumn,showpacs,preprintnumbers,prb]{revtex4-1}

\usepackage{graphicx,epsfig,color}
\usepackage{times}
\usepackage[scr=rsfs,cal=boondox]{mathalfa}

\begin{document}
\title{Spin orbit interaction in nanotubes}
\author{C. A. B\"usser}  
\affiliation{Instituto de Ciencias B\'asicas y Exeperimentales, Universidad de Mor\'on, Buenos Aires, Argentina  }
\email{Corresponding author: busserc@gmail.com}
\begin{abstract}
In recent years, silicene, germanene, and stanene have received considerable attention due to their possibilities to show a spin Hall effect. Nanoribbons made of these materials are expected to have topologically protected states.

In this work, we study the electronic properties of nanotubes made of Si, Ge, Sn, and functionalized Sn. The main difference between these materials and graphene is the relevance of spin-orbit interaction.
The lack of edge states in a seamless tube eliminates the possibility to find a topological edge state. The spin-orbit interaction breaks the degeneracy of Dirac's cones and eliminates the chances of finding a metal nanotube. As a consequence, this transforms all nanotubes with spin-orbit interaction in trivial band insulators.

We focus our attention on two features. First, we study the energy band gap as a function of the diameter of the nanotubes. Then, we concentrate on controlling the band gap of a nanotube by applying an external radial electric field.
\end{abstract}
\maketitle

\section{Introduction}
\label{introduction}
One of the consequences of the reduction in the size of electronic circuits due to Moore's law is the appearance of new effects coming from quantum mechanics. These new effects, which appear on the scale of nanometers, can be used to improve the performance of new circuits in the electronic industry. This new area of knowledge is usually called {\it nanoelectronics}.

In the field of solid-state physics, the appearance of materials associated with carbon, such as {\it graphene}\cite{graphene1,NRB,nanotechbook}, has aroused great interest due to its possible applications in electronics. Graphene, which is a monolayer of carbon atoms with a planar hexagonal structure, is a zero-gap {\it two-dimensional} (2D) semiconductor, {\it i.e.} there is no separation between the conduction and valence bands The lack of gap makes it very difficult to use it to build an effective field effect transistors (FETs) but due to its dielectric properties, graphene combined with silicon or germanium could be used in the electronic components industry.\cite{Chhowalla16}

Graphene is the first experimentally made monolayer material.\cite{graphene2} Its energy band structure near the Fermi level is described by a linear relationship, such as Dirac's relativistic electron theory. That is why the electrons close to the Fermi level of graphene are often referred to as {\it Dirac fermions}. This leads to several novel physical properties, but above all, it generates surprising academic interest. Thus, the success of graphene has triggered an extensive search for other monolayer materials. In particular, monolayer topological materials are very interesting since they can be topological insulators~\cite{kanemele} or topological superconductors\cite{Ezawa15}. It is for all this that it is expected that graphene, and similar compounds, will have a strong impact on the development of components in the electronics industry.

Other materials, which are also 2D semiconductors, would not present the problem of the zero band gap. For example, molybdenum disulfide (MoS$_2$) has a forward gap of 1.8eV.\cite{MoS2} It is feasible to find 2D materials, where charge carriers are constrained to move in two dimensions, that can be used to create new devices taking Moore's law one step further in its size reduction.\cite{Lu07,Chhowalla16,Yu20,Das21}

Other physical systems that are expected to have a strong impact on the development of the electronic components industry are the {\it carbon nanotubes} (CNTs).\cite{saito93,kanemele97,nanotubesbook,nanotechbook} These systems consist of tiny tubes of carbon atoms with diameters between 0.5 and 1 nanometers and a few micrometers in length. Depending on how the CNTs have been wound, they can be metallic or semiconductors. In the case of semiconductor CNTs, if a magnetic field is applied along their longitudinal axis, the insulating gap can be closed, transforming them into metal.\cite{nanotubesbook,saito2000,nanotechbook} 

A natural question is whether there are other systems with properties similar to those of graphene, but with other types of atoms. It has been shown that it is possible to fabricate monolayer systems with a hexagonal structure made of silicon, germanium, and tin, which are called silicene, germanene, and stanene, respectively.\cite{silicene1,Ezawa15}

Germanene and stanene are expected to be topological insulators (TI). A topological insulator is a state of matter characterized by an insulating space in the center accompanied by topologically protected metallic character edges.\cite{kanemele} Frist principles calculation had also predicted that, with a decorated or {\it functionalized} stanene atom (xSn), can be obtained larger insulating energy gaps in the bulk states but keep the metallic edges.\cite{Xu13} Thus, the physics of these materials is located at the confluence of graphene and topological insulators, which results in very interesting physics with many possibilities of also having an impact on the electronics industry.

To study the TI is used a tight-binding model between second neighbors. This tight-binding model with complex coefficients to second neighbors in a hexagonal lattice has already been used by F. D. Haldane to study the quantum Hall effect (QH) in the absence of a magnetic field ({\it i.e.} without Landau levels)\cite{haldane88}. Haldane found that charge transport only occurred at the edge of the system, while the bulk states were insulators (as is the case with the Hall effect). More recently, C. L. Mele and E. J. Kane used this same tight-binding model to represent spin-orbit (SO) coupling in graphene. They found that the ground state of this system exhibits a spin QH effect and has a non-trivial topological order that is robust against small perturbations like lattice disorder.\cite{kanemele} This leads to a system that is topologically distinct from a band insulator. As in the QH in the hexagonal lattice, the electronic transport takes place at the edge of the lattice while the bulk of the 2D structure remains insulating. M. Ezawa has done a complete analysis and classification of these topological insulators.\cite{Ezawa15} In this work was found that ribbons of Ge and Sn are strong candidates to be TI. He also proposed a FET, or a topological quantum transistor, using a nanoribbon of Si.\cite{Ezawa13}

Nanotubes (NT) of these TI materials as germanene or stanene will lead to the elimination of the edge states transforming the tubes into semiconductors. In this work, we center our attention on the electronic properties of nanotubes with a spin-orbit coupling as can be expected in germanene (GeNT) and stanene (SnNT).

Other interesting studies have been focused on the interaction between magnetic impurities and graphene\cite{Fritz13}  or CNT\cite{busser11,busser12a}. In the case of the interaction with TI ribbons, and using an exact mapping\cite{busser13b}, they have found that, even if the topological state is located at the edge of a ribbon, there is a sharp distinction between the Kondo effect ({\it i.e.} a many-body effect) for an impurity located at a crest or a trough site at the zigzag edge.\cite{Allerdt17a} This impurity position dependency has been corroborated using Majorana correlations in topological systems.\cite{diniz23} This shows that to have a correct insight into many-body effects is necessary to understand correctly the electronic properties of these new compounds. For this reason, as we already have stated, we focus our attention on tubes made of Si, Ge, Sn, and functionalized tin (XSnNT).

This work is organized as follows, in Section~\ref{honeycomb} we review the properties of graphene and compare them with silicene, germanene, and stanene. In Section~\ref{results} we present the results for the energy band gap for tubes made of germanium and tin while in Sec.~\ref{transistor} we discuss how to control the energy gap and the possibility to design a field effect transistor using such tubes. Finally, in Sec.~\ref{conclusions} we present our conclusions.

%\centerline{ \rule{5cm}{5pt} }
%====================================================================
\begin{figure}
\centerline{\epsfxsize=3.80cm\epsfbox{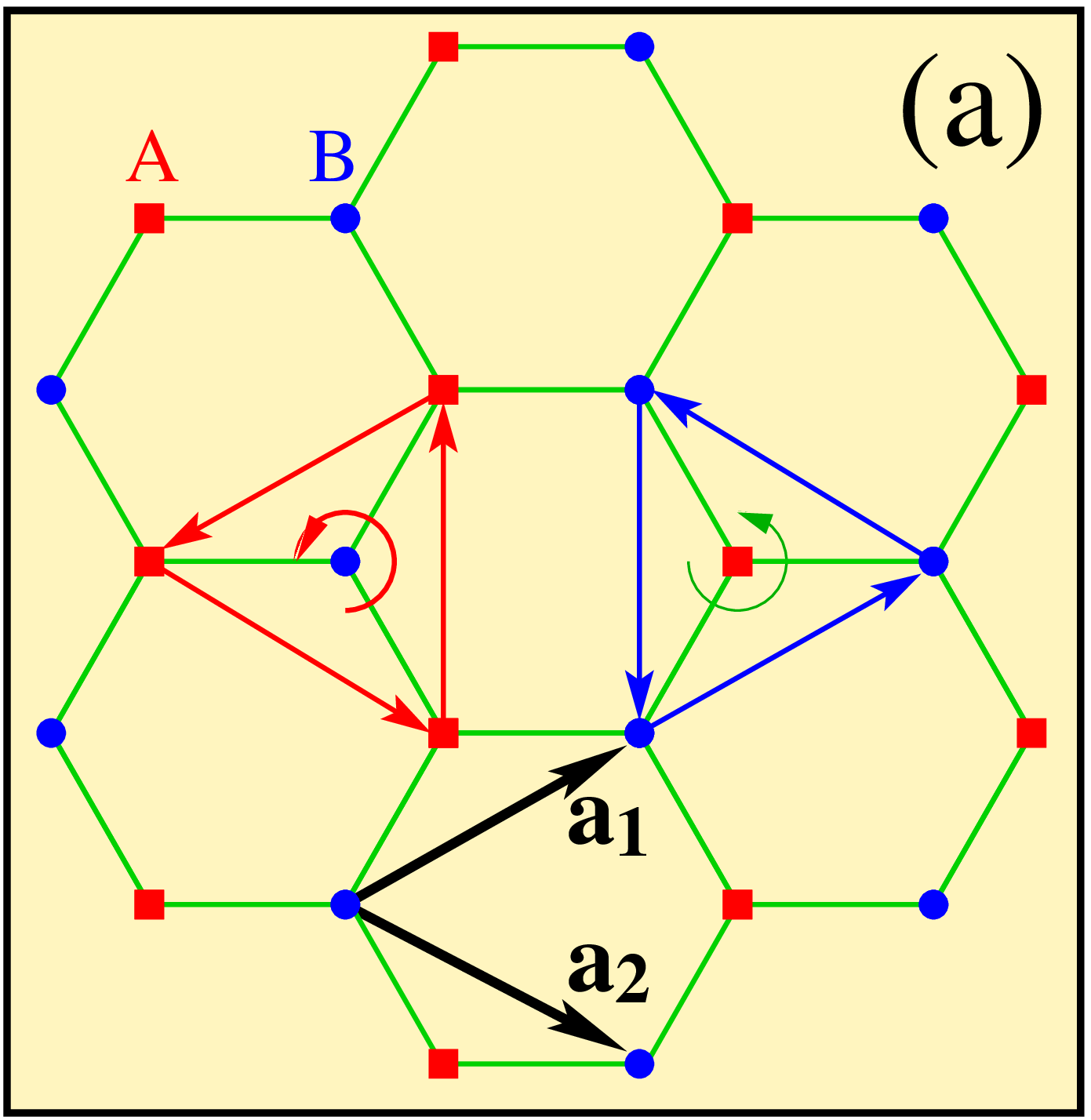}\hspace{0.4cm}\epsfxsize=4.075cm\epsfbox{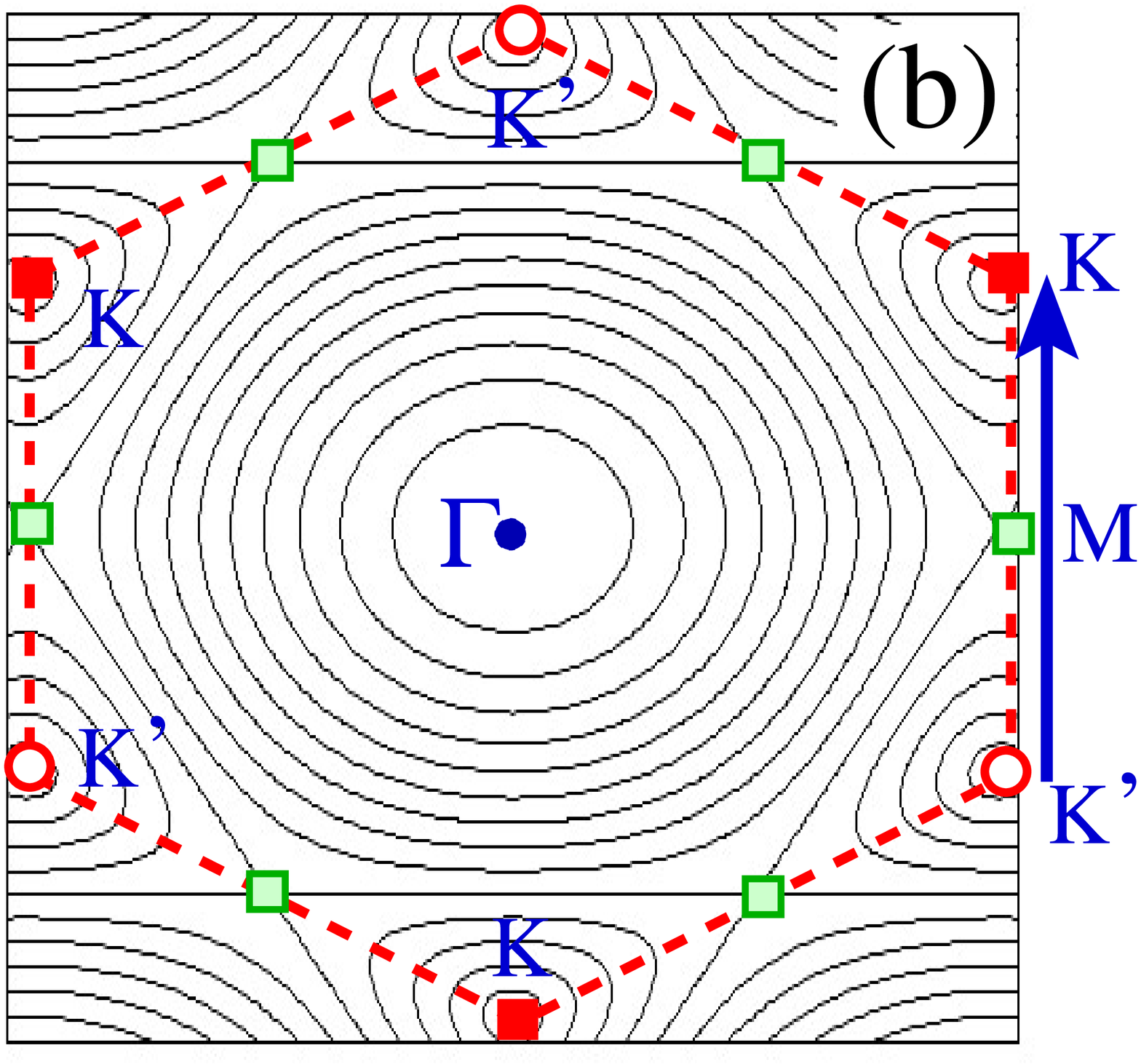}}
\caption{(Color on line) (a) -- schematic picture for the origin of the intrinsic spin-orbit interaction as a second neighbor process. Squares  (red) and circles (blue) represent sublattices A and B respectively. The arrows indicate the closed trajectory of an electron from a carbon atom in sublattice A or B that encircles the spin of the electron in a carbon atom in sublattice B or A respectively. The lattice vectors, $\vec{a}_1$ and $\vec{a}_2$, are indicated. (b) -- Reciprocal lattice of the graphene. The equi-energy curves are shown, they are circles around the K points and the center of the Brillouin zone $\Gamma$. The M points of the boundary of the BZ zone are connected by straight lines. Red dashed lines enclose the Brillouin zone. The straight arrow shows the trajectory in $k$-space used to display the energy dispersion. } 
\label{figure-01}
\end{figure}
%====================================================================
\section{Two dimensional honeycomb lattice with spin-orbit interaction}
\label{honeycomb}
As we have already said, graphene has a hexagonal lattice structure, graphene has a hexagonal lattice structure,  which can be characterized by two interpenetrating triangular sub-lattices. Note that, for graphene, only first-neighbor interactions are considered.

Near the Fermi level, the energy bands of graphene have a linear relationship with the wave vector $k$. As we mentioned, this is a characteristic that this system shares with the relativistic electrons described by the Dirac equation. Hence, the study methods developed for the Dirac equation can be applied to graphene. This special linear relationship between energy and $\vec{k}$ occurs at six points in the reciprocal space. These points are called Dirac's cones (DC). In Figure~\ref{figure-01}(a) we show the reciprocal space indicating the six DC (K and K' points).

One may wonder if other materials have similar properties to the CNT. If we look at the periodic table of elements, we will see that below carbon, in the same column 14, silicon, germanium, and tin can be found. All these elements have a similar last electronic shell ($s^2p^2$) allowing similar chemical bonds. That is why with these three elements it is expected that monolayers with a hexagonal structure can also be formed. Silicene sheets have already been fabricated.\cite{aufray10,lalmi10} Is for this reason that, as we brought up in the Introduction, this honeycomb structure can also be made of Si, Ge, and Sn. Although their lattice structure is similar, a hexagonal structure made up of two interpenetrating triangular sub-lattices, these two triangular lattices lie in different planes separated by a distance $2l$.  This enables interaction with second neighbors in the tight-binding model that gives a band structure different from that of graphene.\cite{Ezawa15}

%====================================================================
\begin{figure}
\centerline{\epsfxsize=7.5cm\epsfbox{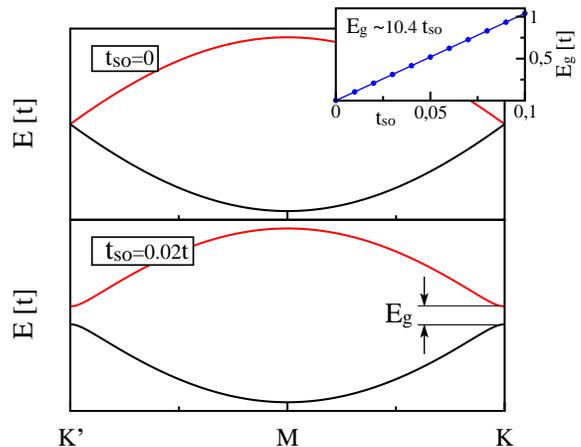}}
\caption{(Color online) Energy dispersion for the graphene $\pi$-bands ($t_{so}=0$) and for a hexagonal lattice with the spin-orbit interactions ($t_{so}=0.02t$). In the inset, we show the gap as a function of $t_{so}$.} 
\label{figure-02}
\end{figure}
%====================================================================

The independent electron Hamiltonian $H_{\rm band}$ that describes the 2D honeycomb lattice for C, Si, Ge, Sn, and functionalized Sn is similar and corresponds to a tight binding band structure,
\begin{eqnarray}
H_{\rm band} &=& H_{\rm hex} ~+~ H_{\rm SO} ~+~ H_{Ez} \label{Hband}\\
 H_{\rm hex} &=& -t     \sum_{\langle i,j\rangle\sigma}   c_{i\sigma}^\dagger ~c_{j\sigma}  \label{Hhex}\\
 H_{\rm SO}  &=&  i~t_{so}\sum_{\langle\langle i,j\rangle\rangle\sigma} \sigma \nu_{ij} c_{i\sigma}^\dagger ~c_{j\sigma}  \label{Hso}\\
 H_{Ez}      &=& E_0    \sum_{i\sigma}    \mu_i   c_{i\sigma}^\dagger ~c_{i\sigma}  \label{HEz}
\end{eqnarray}
where $c_{i\sigma}^\dagger$ and $c_{i\sigma}$ are the usual operators that create and destroy an electron at site $i$ with spin $\sigma$. The spin index $\sigma$ means $\sigma=\uparrow,\downarrow$ when acts as a subindex and has a value $\sigma=\pm 1$ when is inside an equation. The parameter $\mu_i$ is to differentiate the non-equivalent sites of the hexagonal lattice; it has the value $\mu_i=1$ when $i$ points to a site A and $\mu_i=-1$ when it points to a B site. Sums over $\langle i,j\rangle$ and $\langle\langle i,j\rangle\rangle$ runs over the first or second neighbors respectively.

In this Hamiltonian, $H_{\rm band}$, we considered three terms. The first term, $H_{\rm hex}$, represents the usual nearest-neighbor hopping with the transfer energy $t$ and takes into account the tight-binding model of the honeycomb lattice. In Figure~\ref{figure-01}(b) we have schematized the interactions with the first neighbors (green lines). The second term, $H_{\rm so}$, represents the effective SO coupling with $t_{so}$. Note that, as the terms in $H_{\rm so}$ have a complex phase, one must be careful with the direction of circulation (indicated by arrows in the figure). This is taken into account by setting the parameter $\nu$. In this case, $\nu_{ij}=+1$ if the next-nearest-neighboring hopping is anticlockwise and $\nu_{ij}=-1$ if it is clockwise respect to the positive $z$-axis.  Let us observe that the second neighbor terms connect sites of the same triangular sublattice A or B. The direction of circulation around each site of the lattice has also been indicated in the figure.

Finally, the third term in Equation~\ref{Hband}, $H_{Ez}$ represents the staggered sublattice potential when a perpendicular external field $E_0$ is applied. One of the characteristics of Si, Ge, and Sn is the buckled structure that allows the application of different external fields between the sublattices A and B. As a consequence, $E_0$ can tune the band gap when the SO interaction is present. In Section~\ref{transistor} we will come back to this term.

\begin{table}[]
{{%\large 
\begin{tabular}{|l||c|c|c|c|}
\hline
 lattice   & ~t~[eV]~ &  ~$t_{so}/t$~ & ~$a_0$~[nm]~  \\
 \hline
 \hline   
 graphene  & 2.8    &  $\sim 2~ 10^{-5}$  &  0.246 \\
 silicene  & 1.6    &      0.0005      &  0.386 \\
 germanene & 1.3    &      0.0064      &  0.402 \\
 stanene   & 1.3    &      0.015       &  0.470 \\
 F-stanene & 1.3    &      0.044       &  0.470 \\
 \hline
\end{tabular}  }}
\caption{In this table, we present the parameters that characterize graphene, silicene, germanene, and stanene. The parameters $t$ and $t_{so}$ are the coefficients of the tight-binding model to first and second neighbors respectively. $a_0$ is the lattice parameter. }
\label{tabla1}
\end{table}

Table~\ref{tabla1} indicates the relationship between the first neighbor term ($t$) and the second neighbor term ($t_{so}$) in the tight-binding model for graphene, silicene, germanene, and stanene calculated by first principles.\cite{Ezawa15,Allerdt17a} We see that $t_{so}$ is negligible for the case of graphene, but begins to be significant for the other systems. Along this work, unless other this is indicated, we will take the hopping parameter $t$ as the unit of energy. Table~\ref{tabla1} is then relevant to understand the relationship between C, Si, Ge, and Sn.

In Figure~\ref{figure-01}(b) we show a representation of the reciprocal $k$-space. For a hexagonal lattice, the reciprocal space is also a hexagonal lattice rotated in $\theta=\pi/6$. The first Brillouin zone is represented by the hexagonal dashed lines. Points $\Gamma$, K, and M of the Brillouin zone are indicated. Observe that as the reciprocal space is also a hexagonal lattice the K points are also formed by two interpenetrated triangular lattices. This defines two non-equivalents K and K' points, where we can find the DCs. As the $\Gamma$ point of the reciprocal space is almost not affected by the SO interaction we will focus our attention on the trajectory that joins the two different DCs points, K and K' passing through the point M. This is indicated in the panel~(b) by the blue arrow.

In Figure~\ref{figure-02} we show the energy dispersion of the Hamiltonian given by Eq.~\ref{Hband} calculated in the $\vec{k}$ direction schematized by the blue arrow of Fig.~\ref{figure-01}(b).\cite{nanotubesbook}  We present two cases in absence of the external field. In the upper panel, we present the case of $t_{so}=0$ that corresponds to the graphene. We can see the linear relationship of Dirac's cones around the K and K' points. In the lower panel, with $t_{so}=0.02$, we present the result for a case where the SO interaction can not be neglected. A gap ($E_g$) split the Dirac's cones for all K and K' points. In the inset, we can observe the $E_g$ as a function of the strength of the SO interaction $t_{so}$. This relation between $E_g$ and $t_{so}$ is linear, roughly $E_g=10.4t_{so}$, as expected by previous work.\cite{Ezawa15} Observe that this band structure is independent of the spin projection.

In the next section, we will study the effect of the SO interaction in nanotubes.

%====================================================================
\begin{figure}
\centerline{\epsfxsize=8.5cm\epsfbox{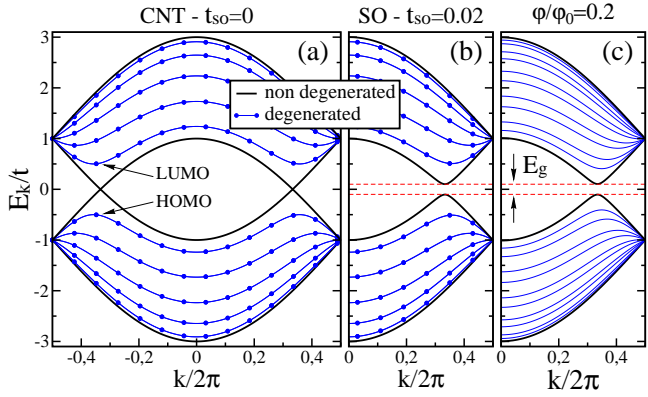}}
\caption{(Color online) Energy bands for a (6,6) metallic armchair nanotube. 
(a) Energy bands for the CNT ($t_{so}=0$). Degenerate states are indicated by dotted lines. The band crossing, for armchair NT, occurs at a value of  $k =\pm{2\pi}/{3}$. The system is metallic. In this panel, we have indicated the HOMO ($v_1$) and LUMO ($c_1$) points to compare with this molecular orbital energy gap.
(b) Spin-orbit coupling ($t_{so}=0.02$) opens a gap at $k =\pm{2\pi}/{3}$ but keep all degeneracies.
(c) Same qualitative behavior is observed if an external magnetic flux $\varphi$ is included with $t_{so}=0$. In this late case is broken all the degenerated states of this armchair NT.}
\label{figure-03}
\end{figure}
%====================================================================
\section{Nano tubes with and hexagonal lattice and spin-orbit interaction}
\label{results}
A nanotube can be described as a single layer of a crystal that is rolled up into a seamless cylinder, one atom thick. Usually, it has a few tens of atoms along its circumference and a length of several micrometers along the axis of the cylinder\cite{nanotubesbook}. This nanotube is described by the chiral vector $C_h$, which specifies the shape of the base of the nanotube,
\begin{equation}
\vec{C_h} = m \vec{a_1}+n\vec{a_2}  ~\equiv~ (m,n),
\label{eq_Ch}
\end{equation}
where $\vec{a_1}$ and $\vec{a_2}$ are the translation vectors of one of the triangular sublattices.\cite{nota-a0} This two vectors are indicates in Fig.~\ref{figure-01}(a). In this way, the vector $\vec{C_h}$ points to two equivalent sites. Nanotubes can be fully identified by their chiral numbers $(m,n)$. Many of its physical properties, such as its thickness or the number of atoms in the unit cell, are fixed by these two numbers. For example the diameter $d$ of a NT is calculated after the modulus of $\vec{C_h}$,
\begin{equation}
d = a_0 \sqrt{(m+n)^2-mn}/\pi,
\end{equation}
where $a_0$ is the lattice parameter shown in table~\ref{tabla1}. Observe that $\vec{C}_h$ is the circunference of the tube.
 
About 1/3 of the CNTs are metallic, while the rest are semiconductors. It can be shown that when the number $q$
\begin{equation}
q=\frac{2n+m}{3},
\label{eq_q}
\end{equation}
is an integer, the CNT will be metallic.\cite{dresselhausbook98} It is easy to see that all nanotubes with chiral numbers with $m=n$ will be metallic. These nanotubes, with repeated chiral numbers (m,m), are called {\it armchair} due to the shape of its unitary cell ({\it i.e.} the base of the tube).

%====================================================================
\begin{figure}
\centerline{\epsfxsize=8.0cm\epsfbox{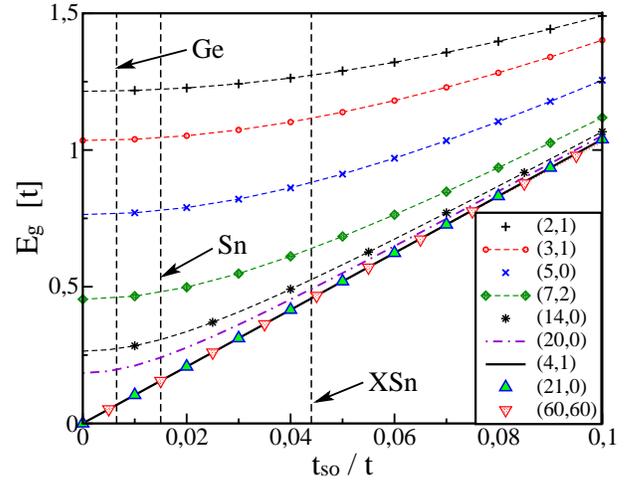}}
\caption{(Color online) We can see the energy gap ($E_g$) for different nanotubes as a function of the SO interaction $t_{so}$. The specific $t_{so}$ for nanotubes of germanene, stanene, and functionalized statene (XSn) are indicated with vertical dashed lines. } 
\label{figure-04}
\end{figure}
%==================================================================== 

We performed tight-binding calculations with a Hamiltonian that included the SO interaction to obtain the changes in the band structure introduced by this term. The energy bands are calculated as a function of $k$, with $k$ being the momentum along the axis of the tube. In Figure~\ref{figure-03}(a) we present the bands structure of an armchair CNT ($t_{so}=0$) with chiral numbres (6,6). In this case, the band structure does not present an energy band gap and is clearly metallic. For all the armchair CNT the band crosses at $k=\pm 2\pi/3$. These two points represent the two non-equivalent K and K' points. Compare the none degenerated bands with the energy dispersion shown in Fig.~\ref{figure-02}(a) ($t_{so}=0$). An alternative way to calculate the NT energy dispersion is by slicing the 2D bands.\cite{saito92,saito93} Can be shown that for the armchair tubes, one of the slices is in the direction of K'$\to$M$\to$K. It is this slice that gives the inner metallic band. \cite{saito2000} 
%%TO CANCEL
%%See more information in the supplementary material.

We have also indicated in Fig.~\ref{figure-03}(a) the highest occupied (HOMO) and the lowest unoccupied (LUMO) molecular orbitals of the bulk states. We will compare the HOMO-LUMO gap with the split of the Dirac's cones.

As we have shown in the previous section, the spin-orbit interaction split the DC for the 2D systems. In such a case, as the 2D system is not metallic, the formed tube is a semiconductor. We have to note that all the tubes are at half-filling. Here lies the bigger difference between the NTs and the nanoribbons with SO interactions. In the case of nanoribbons, the edge states are metallic and topologically protected. For construction, the NT has no edges, and as a consequence, there are no metallic states. All the NTs with SO interaction are, therefore, semiconductors. Panel~(b) of Fig.~\ref{figure-03} presents a clear example of this issue. We can observe the energy band structure for a (6,6) NT with $t_{so}=0.02t$. As discussed above, the bands with linear dispersion are the only non-degenerate ones. The SO interaction does not break this degeneracy. The split of the  Dirac's cones opens a gap $E_g$ proportional to $t_{so}$. As before $E_g=10.4 t_{so}$.

We also calculated, within the tight-binding approximation and in the absence of the SO interaction, the changes in the band structure produced by an external magnetic field along the principal axis of the NT. We do not consider a magnetic field large enough to turn on a Zeeman effect, so both spin populations remain the same. The outcome of this magnetic flux is shown in panel~(c), we can observe that the opened gap is proportional to the magnetic flux $\varphi$.\cite{nanotubesbook} Observe also, that all the degenerated bands, for the armchair NTs, split. We can compare the effects of the intrinsic spin-orbit interaction with the ones produced by an external magnetic field along the principal axis of the nanotube. Note, however, that although both interactions, the SO and the magnetic flux, open a gap in the band structure, the energy gap produced by the intrinsic spin-orbit interaction only affects the metallic band.

%====================================================================
\begin{figure}
\centerline{\epsfxsize=7.5cm\epsfbox{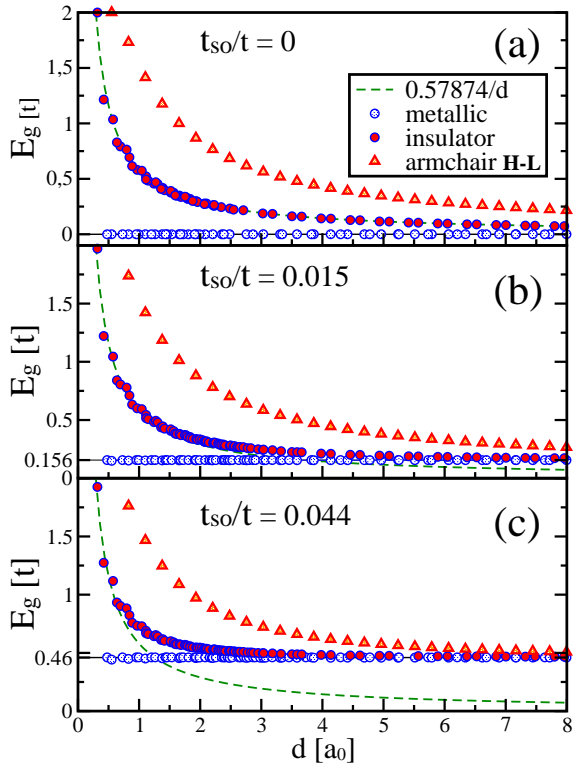}}
\caption{(Color online) Energy gap as a function of NTs diameter $d$ in units of the lattice parameter $a_0$. In panel~(a) we can see the band gap, $E_g$, for the metallic and insulators tubes for the CNT case as a function of the tube diameter $d$. 
In panels~(b) and~(c) we can show the gap $E_g$ for the metallic and insulators tubes for the SnNTs ($t_{so}=0.015$) and functionalized XSnNTs ($t_{so}=0.044$) cases as a function of $d$. A dashed line in all panels shows the fitted curve for the gap dependency for semiconductors CNTs. }
\label{figure-05}
\end{figure}
%====================================================================

%%% For Figure 4
To understand the effect of the SO interaction in NTs we have calculated the values of $E_g$ for tubes with different chiral numbers (m,n) as a function of $t_{so}$. These results are presented in Figure~\ref{figure-04}. For $t_{so}=0$ the tubes that have an integer $q$ are metallic, but as soon as $t_{so}$ increases an insulating gap appears and all the NTs became semiconductors. Vertical dashed lines indicate the $t_{so}$ for GeNTs, SnNTs, and XSnNTs. The insulating gap of the nonmetallic NTs also increases as the split of the DCs pushes up all the states of the 2D systems that support the density of states of the NTs. For this reason, two different NTs, (21,0) and (20,0), (one being metallic and the other with a relatively large insulating $E_g$ of 0.55~eV for a CNT) have a very similar gap when they are made of functionalized Sn. 
%%TO CANCEL
%%In the supplementary material we present and discuss, as an example, the band structure for these two tubes.

%%% For Figure 5
To finish this section,  we study how $E_g$ behaves as a function of the diameter of the NT. We calculated within the tight-binding model, the gap for all chiral numbers (m,n) with m from~$2$ to~$15$ and $n$ from zero to $m$. Had been reported, for the semiconductor CNT a proportional inverse relationship between $E_g$ and $d$.\cite{kanemele97,saito2000} This is shown in Figure~\ref{figure-05}(a) where the metallic tubes ({\it i.e.} $E_g=0$) are indicated with open circles while the semiconducting tubes are shown with full circles. We found that $E_g\sim 0.57/d$ with the gap in units of the hopping parameter $t$ and the diameter in units of  $a_0$. The fitting curve is shown with dashed lines in the figure.\cite{saito93} We also present the HOMO-LUMO gap for the metallic armchair CNT. This HOMO-LUMO gap follows a different scaling with the diameter of the nanotube.\cite{kanemele97} We will use this gap as a reference to compare with what happens when the SO interaction is applied.

%====================================================================
\begin{figure}
\centerline{\epsfxsize=8.5cm\epsfbox{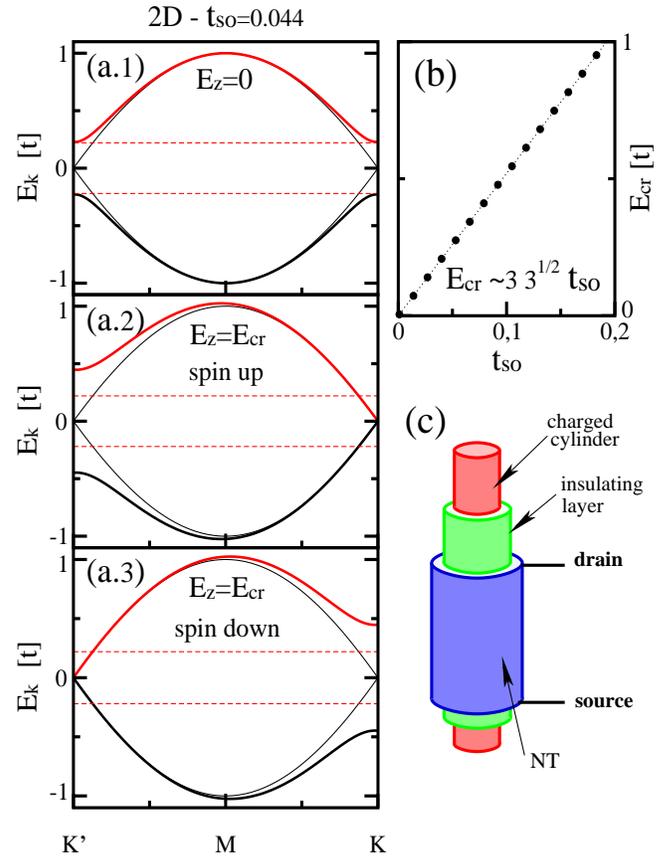}}
\caption{(Color online). In panel (a) we present the difference in the 2D system when a perpendicular electric field $E_0$ is applied. In panel (a.1), with thick lines, we can observe the band structure in the direction K'$\to$M$\to$K when the SO interaction is considered. With thin lines, we show, to compare, the band for $t_{so}=0$. In the lower panels (a.2) and (a.3) we see the effect on spins when the electric field $E_0=E_{cr}$ that closes the gap is achieved. Observe that the gap closes for different K points.
Panel (b) show the relationship between the critic field $E_{cr}$ as a function of $t_{so}$.
Finally, panel (c) shows a schematic of the device to achieve an electric field perpendicular to the surface of an NT.}
\label{figure-06}
\end{figure}
%==================================================================== 

The effect of the SO interaction is shown in Figure~\ref{figure-05} panels~(b) and~(c). We have chosen the parameter $t_{so}$ that describes the SnNTs and the functionalized XSnNTs as they present the strongest effects. In both cases, we observe that the tubes that were metallic for the case of CNTs, with an integer number of $q$ in Eq.~\ref{eq_q}, are now semiconductors. All these tubes have the same gap. We found $E_g=0.156t$ (or 0.2eV) for the SnNts and $E_g=0.457t$ (or 0.59eV) for the XSnNTs. These values for $E_g$ are in agreement with the split of the DC found in the previous section for the 2D system.

For the chiral number that corresponds to the semiconductors CNTs, those with a fractional value of $q$, we found that the gap for the small diameter tubes is following approximately the same decaying relation found for the CNTs (dashed line). For large diameter $d$ however, we found that the gap goes to the asymptotic limit of the $E_g$ found for the metallic tubes. This can be understood as there are no states as a consequence of the split of the DC in the 2D system that can be used as a base to construct a metallic tube.

We can observe also that the HOMO-LUMO gap has a similar behavior. Thus remains the same as the CNTs for small diameter tubes and achieves the value of the split of the Dirac's cones for large diameter. The split of the DC displaces all the states including the HOMO and LUMO points.

\begin{table}[]
{{%\large 
\begin{tabular}{|l||c|c|c|c|}
\hline
 lattice   &   ~band width~[eV]~ & ~2Ds$~E_g$~~[eV]~ & ~$E_{cr}$~~[eV]~  \\
 \hline
 \hline   
 CNT    & 16.8   & $6~10^{-4}$ & $\sim 10^{-3}$\\
 SiNT   & 9.6    &  0.008      &  0.003 \\
 GeNT   & 7.8    &  0.08       &  0.04 \\
 SnNT   & 7.8    &  0.20       &  0.10  \\
 XSnNT  & 7.8    &  0.59       &  0.29  \\
 \hline
\end{tabular}  }}
\caption{In this table, we present the bandwidth, the energy gap, and the critical field $E_{cr}$ that characterize graphene, silicene, germanene, stanene, and functionalized statene NTs. They are expressed in eV. }
\label{tabla2}
\end{table}

\section{Controlling the energy gap in nanotubes}\label{transistor}

In the previous section, we analyzed the effect of the SO interaction in the nanotubes. We found that all the NTs with $t_{so}>0$ are semiconductors. Now we will analyze the effect of an external electric field.  In the tight-binding model, this effect is included in the Hamiltonian by the term $H_{Ez}$ given by Equation~\ref{HEz}. 
As mentioned in Section~\ref{honeycomb}, the honeycomb lattice for the Si, Ge, and Sn is distorted and forms a bucked structure. The two sublattices, A and B, are separated by a distance $2\mathcal{l}$.
Then this structure generates a staggered sublattice potential under the external field $E_z$ perpendicular to the surface. The parameter in Eq.~\ref{HEz} became an effective external field,
\begin{equation}
E_0=\mathcal{l}~E_z.
\end{equation}

As we already mentioned, and following References~\onlinecite{Ezawa15} and~\onlinecite{Ezawa12}, we start this discussion by analyzing the 2D system. 

In Figure~\ref{figure-06}(a.1) we present the effect of applying the external field $E_0$. The upper panel shows the band structure for $t_{so}=0.044$  that correspond to the functionalized Sn. As described before, the degeneracy at the unequal reciprocal points K and K' splits when the SO interaction is considered. The system in the bulk 2D became a band insulator.\cite{bandKKp}

Is important to note that this gap $E_g$ is tunable by controlling the external field $E_0$. When this field is applied the insulating gap became narrower.\cite{Ezawa12} At a critical field, $E_{cr}$ the gap is completely closed and the 2D system became metal. In panels (a.2) and (a.3) we present this situation for both spins projection. For spin-up (a.2) we can observe that is gapless for the K point while having a gap $2E_g$ for the reciprocal point K'. On the contrary, spin-down (a.3) is gapless for the K' point and has a gap of $2E_g$ at K.

Once again, in this effort, we are not interested in the edge states that can transform the honeycomb structure in topological insulators. Rather, we are focused on the bulk states that will provide support to the band structure of the NTs with spin-orbit interactions.

Finally, for an external field $E_0$ larger than $E_{cr}$ the gaps open again for all the closed Dirac's cones.

%====================================================================
\begin{figure}
\centerline{\epsfxsize=7.5cm\epsfbox{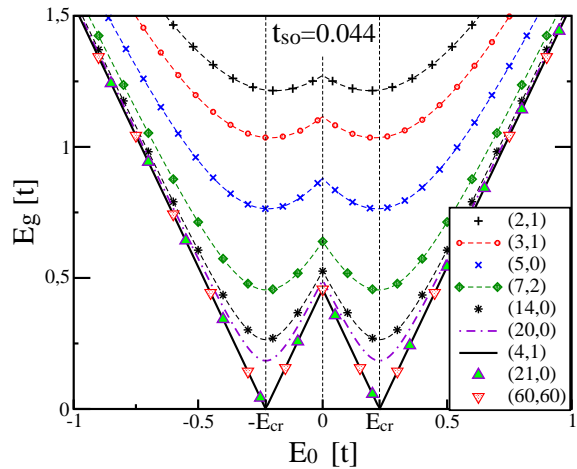}}
\caption{(Color online). Energy gap ($E_g$) for different nanotubes as function of the radial electric field $E_{0}$ perpendicular to the NT surface. For these results we have used $t_{so}=0.044t$ that represent the case of the XSnNT.
All the metallic tubes shown, (4,1), (21,0) and (60,60), present the same behavior since it depends on the split of the Dirac's cones. For the metallic NTs the gap is closed for $\pm E_{cr}$. }
\label{figure-07}
\end{figure}
%====================================================================

In Figure~\ref{figure-06}(b) is shown the critical field $E_{cr}$ as function of $t_{so}$. This critical field can be calculated as,\cite{Ezawa12,Ezawa15}
\begin{equation}
E_{cr}=3\sqrt{3}~t_{so}
\end{equation}

Observe that the electric field $E_z$ is perpendicular to the surface of the honeycomb lattice. To obtain such a perpendicular field we can wrap the honeycomb around an electrically charged cylinder. In this way, the electric field applied to the NT will have the desired symmetry. This device is schematized in Figure~\ref{figure-06}(c).

Now we can analyze the effect of the external field $E_0$ in nanotubes. We start presenting, in Figure~\ref{figure-07}, the insulating gap as a function of $E_0$ for the SO interaction $t_{so}=0.044$ that describes the XSnNTs. We chose the same chiral numbers used in Fig.~\ref{figure-04}. For the chiral numbers that are not metallic CNTs (fractional $q$) we have, for $E_0=0$, the gap produced by the SO interaction discussed in Sec.\ref{results} ($t_{so}=0.044$ in Fig.~\ref{figure-04} corresponding to XSn). 

As soon as we increase the value of $E_0$ we can see that the value of the gap decreases. Then, when $E_0=\pm E_{cr}$, the gap $E_g$ reaches the value it has for CNTs ($t_{so}=0$ in Fig.~\ref{figure-04}). For $E_0>E_{cr}$ the gap opens again.

On the other hand, we can see that the chiral numbers for metal CNTs all exhibit exactly the same behavior. The gap $E_g$ reduces its value linearly with $|E_0|$ up to become zero for $E_0=\pm E_{cr}$ and, beyond this point, opens linear again. This reduction of the gap, when $|E_0|$ goes from zero to $E_{cr}$, can be understood as there are three of the Dirac's cones, depending on the spin, that close for such external field $E_{cr}$.

%====================================================================
\begin{figure}
\centerline{\epsfxsize=7.5cm\epsfbox{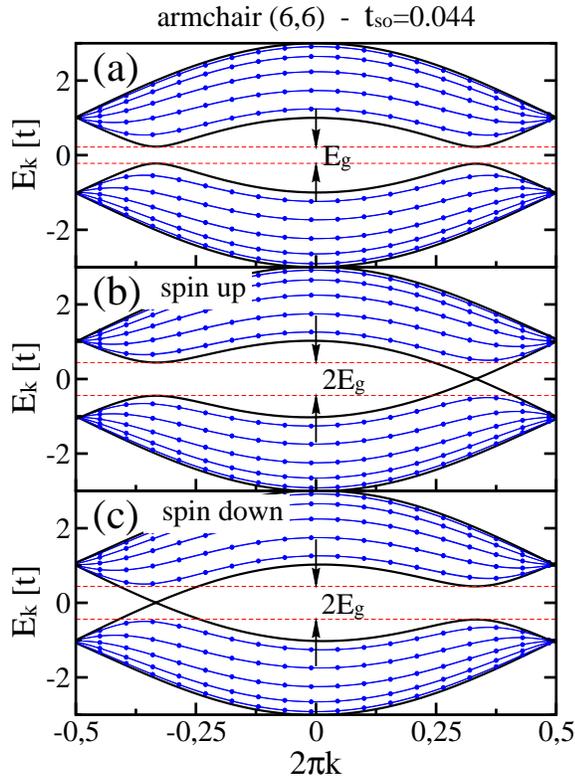}}
\caption{(Color online). Energy band structure for a (6,6) with SO interaction and an external field $E_0$. In absence of SO interaction this NT is metallic. In panel (a) we can observe that when the spin-orbit is considered, in this case, $t_{so}=0.44$  a gap opens and the system is a trivial insulator as discussed in Sec.~\ref{honeycomb}. In panels~(b) and~(c) the effect of applying an external field $E_0=E_{cr}$ is shown. For the spin-up(down) the gap closes for $k=-2\pi/3$ ($k=2\pi/3$) and double for $k=2\pi/3$ ($k=-2\pi/3$. The up and down spins flows into oppositive directions).  }
\label{figure-08}
\end{figure}
%====================================================================
%%%% Figure 8
To understand what happens with the different spins, we analyze the band structure of the armchair tubes as they are related to both unequal K and K' reciprocal points. In Figure~\ref{figure-08}, we present the band structure for an armchair NT with chiral numbers (6,6) and $t_{so}=0.044$. In panel~(a), we observe that a gap $E_g$ appears at $k=\pm 2\pi/3$. As discussed in the previous section,  the split of the DC due to the SO interaction is the origin of this gap. As was already observed the effect of the external field $E_0$ is different for up and down spins. For spin-up, a small $E_0$ reduces the gap at the K' point and opens it more at the K point. On the other hand, it has the opposite effect on the spin-down. In panels~(b) and~(c) we can observe the band structure for a field $E_0=E_{cr}$ that had closed the energy gap of the DC for the 2D system. At this critical field, the gap is zero for spin up at $k=2\pi/3$ (K point) while is $2E_g$ for $k=-2\pi/3$ (K' point). For spin-down, it closes for $k=-2\pi/3$ and is $2E_g$ for $k=2\pi/3$. Then, for the armchair tubes, the up and down spins flow in opposite directions. As a consequence, there is a null net charge current and a pure spin current.

Observe that for $E_0=E_{cr}$ the behavior of the spins is oppositive; {\it e.g.} spin-up close its gap for $k=-2\pi/3$ and double for $k=2\pi/3$.

A similar feature can be found in nanoribbons.\cite{Ezawa13} However there are two important differences with the NTs. First, when the tube is wrapped the edge states of the nanoribbons are eliminated. For the NT all the sites are equal. Then, the spin current is not located at specific sites but all along the tube. There are no special topological states. The second difference for the NT is that the spin current just can be achieved for a specific field $E_0=\pm E_{cr}$. For any other external field, the NT is a trivial insulator.

In Figure~\ref{figure-08} we presented results just for the armchair nanotubes. In the case of semiconducting tubes, {\it i. e.} the one with a fractional number $q$ given by Equation~\ref{eq_q}, we have seen from Fig.~\ref{figure-07}, that there are no external field $E_0$ that close the gap.

The zigzag tubes, with chiral numbers (m,0) and integer $q$, present a different behavior than the armchairs. As with all the nanotubes, a gap appears when the SO interaction is considered. However, for the zigzag nanotubes, both Dirac's cones lie at $k=0$ and not in $\pm 2\pi/3$. For the external field $E_0=E_{cr}$ one of the DC closes its gap while the other doubles it. As both DCs lie at $k=0$ the band structure is the same for the spin up and down.

For the chiral nanotubes (m,n) with $m\ne n$, $n\le 1$, and integer values of $q$, we found three different behaviors. Some NTs behave as armchairs with a K point at $k=2\pi/3$ and K' point at $k=-2\pi/3$. Other tubes behave similarly to zigzags and have both DC at $k=0$. However, there is a third class of tubes that behaves in a similar way to the armchairs but where K points lie at $k=-2\pi/3$ and K' lie at $k=2\pi/3$ inverting the behavior of the spins in the band structure.  
%%TO CANCEL
%%We present examples of these three different band structures in the supplementary information.

%%%%%%%%%%%%%%%%%%%%%%%%%%%%%%%%%%%%%%%%%%%%%%%%%%%%%%%%%%%%%%%%%%%%%%%%%%%%%%%%%%%%%%%%%5
%%%%%%%%%%%%%%%%%%%%%%%%%%%%%%%%%%%%%%%%%%%%%%%%%%%%%%%%%%%%%%%%%%%%%%%%%%%%%%%%%%%%%%%%%5
%%%%%%%%%%%%%%%%%%%%%%%%%%%%%%%%%%%%%%%%%%%%%%%%%%%%%%%%%%%%%%%%%%%%%%%%%%%%%%%%%%%%%%%%%5
%\centerline{ \rule{5cm}{4pt} }
At this point, we want to remark that this device, an NTs wrap around a charged cylinder, can work as a field effect spin transistor. Effectively, by applying a gate potential to the central cylinder we can change its charge and the value of $E_0$. Thus, we can control the insulating gap transforming the device from a semiconductor to a metal for $E_0=E_{cr}$. Now, if $|E_0|=E_{cr}$, applying a small source-drain potential less than $2E_g$ a spin current will be induced. A similar device had been proposed by M. Ezawa using Si nanoribbons taking advantage of the topologically protected edge states.\cite{Ezawa13} In this case, the transistor acts as a topological insulator if $|E_0|<E_{cr}$ and a band insulator for $|E_0|>E_{cr}$.

In table~\ref{tabla2} we summarize the information on the bandwidth of the NTs, the gap $E_g$, and the external critical field $E_{cr}$ that must be applied to close the gap when the SO interaction is applied. This information is presented for carbon, silicon, germanium, tin, and functionalized tin. The parameters $E_g$ and $E_{cr}$ are valid for the nanotubes with integer values of $q$, {\it i. e.} the ones that are metallic when the SO interaction is not present (CNTs).

Observe that we present $E_{cr}$ in eV and not in the regular units for the electric field of eV/m. This parameter is directly the gate potential that must be applied to the internal cylinder to control the band gap of the proposed transistor.

\section{Conclusions}\label{conclusions}
In this work, we have analyzed the effect of spin-orbit interaction in rolled cylinders or nanotubes. Note that this SO interaction is the origin of the edge states of topological insulators in nanoribbons. When a nanoribbon is rolled-up into a seamless tube, edge states are eliminated. Thus, only the insulating states of the ribbon remain. For this reason, all nanotubes that were metallic for CNTs now became trivial band insulators.

We have analyzed the behavior of the insulating gaps as a function of the diameter of the nanotubes. We found that NTs with the chiral number corresponding to a metallic one for CNTs have a band gap, $E_g$, that does not depend on the radius of the tube. This band gap depends on the strength of the spin-orbit interaction. For the chiral numbers of the semiconducting CNTs we found that, with spin-orbit interaction, they behave as $\sim 1/d$ for small-diameter tubes. However, asymptotically, for large-diameter, they tend to have the same band gap, $E_g$, as the metallic chiral number.

In this effort, we have proposed a device where an external electric field is applied perpendicularly to the surface of the cylinder.
Applying this external field, the band gap $E_g$ can be controlled. At a critical field, $E_{cr}$, the gap is closed and the NTs go from a trivial insulating band to a metallic system. We found that in some cases, with this field $E_{cr}$, the electrons with different spins move in oppositive directions.

Controlling the electric field using a gate potential this device can work as a field effect transistor, where the conductance can be modified with this gate.

%%%%%%%%%%%%%%%%%%%%%%%%%%%%%%%%%%%%%%%%%%%%%%%%%%%%%%%%%%%%%%%%%%%%%%%%%
{\bf \it Acknowledgments - } The author is grateful to G. B. Martins for many discussions on the subject and suggests the direction of this effort. 
This work has been supported by the  {\it SeCyT-Rectorado of the University of Mor\'on} under grants { PIO 2020 - 800 201903 00033 UM and PIO 2021 - 800 202101 00033 UM}.

%###################################################################################
%###################################################################################
\bibliographystyle{apsrev4-1}
\bibliography{my_papers,notes,references}

\begin{thebibliography}{32}
\bibitem [{not()}]{nota-a0}%
  \BibitemOpen
  \href@noop {} {\ }\bibinfo {note} {Observe that in the way we have choose the
  lattice vectors we have that $|\vec{a_1}|=|\vec{a_2}|=a_0$, with $a_0$ being
  the smaller distance between two equivalent sites.}\BibitemShut {Stop}%
\bibitem [{ban()}]{bandKKp}%
  \BibitemOpen
  \href@noop {} {\ }\bibinfo {note} {In the band structure we neglect the
  topological edge states as they are not present in nanotubes.}\BibitemShut
  {Stop}%
\end{thebibliography}
%##########################################################################################
%##########################################################################################
%##########################################################################################

%%%%%%%%%%%%%%%%%%%%%%%%%%%%%%%%%%%%%%%%%%%%%%%%%%%%%%%%%%%%%%%%%%%%%%%%%%%%%%%%%%%%%%%%%5
%%%%%%%%%%%%%%%%%%%%%%%%%%%%%%%%%%%%%%%%%%%%%%%%%%%%%%%%%%%%%%%%%%%%%%%%%%%%%%%%%%%%%%%%%5
%%%%%%%%%%%%%%%%%%%%%%%%%%%%%%%%%%%%%%%%%%%%%%%%%%%%%%%%%%%%%%%%%%%%%%%%%%%%%%%%%%%%%%%%%5

\end{document}